# Guiding Testing Activities by Predicting Defect-prone Parts Using Product and Inspection Metrics

Frank Elberzhager,
Stephan Kremer
Fraunhofer IESE
Kaiserslautern, Germany
{first.last}@iese.fraunhofer.de

Jürgen Münch
University of Helsinki
Helsinki, Finland
juergen.muench@cs.helsinki.fi

Danilo Assmann
Vector Informatik GmbH
Stuttgart, Germany
Danilo.Assmann@vector.com

*Abstract*—Product metrics, such as size or complexity, are often used to identify defect-prone parts or to focus quality assurance activities. In contrast, quality information that is available early, such as information provided by inspections, is usually not used. Currently, only little experience is documented in the literature on whether data from early defect detection activities can support the identification of defect-prone parts later in the development process. This article compares selected product and inspection metrics commonly used to predict defect-prone parts. Based on initial experience from two case studies performed in different environments, the suitability of different metrics for predicting defect-prone parts is illustrated. These studies revealed that inspection defect data seems to be a suitable predictor, and a combination of certain inspection and product metrics led to the best prioritizations in our contexts.

## I. INTRODUCTION

Software and software systems, for example mobile phones, cars, or medical devices, are part of everyone's life. Such systems continuously increase in their size and complexity. Consequently, the risks of failures that might lead to serious consequences also increase. In order to develop high-quality software products, a large number of different analytic quality assurance techniques exist, such as different inspection and testing techniques.

However, the costs and effort required for applying them often exceed or consume a large amount of the available budget, especially for testing [1]. Therefore, approaches for predicting defect-prone parts during development are one way to address challenges such as how to reduce effort and improve effectiveness or efficiency. In order to focus quality assurance activities, product and process metrics are often considered, such as size or complexity metrics, or historical defect data. Based, for example, on the assumption that large code modules contain more defects, the available effort can be allocated in a more suitable manner to large code modules. However, defect data from inspections or reviews that are available early are often not considered for such prioritizations, i.e., synergy effects between early and later quality assurance activities are usually not exploited [2].

Therefore, we propose the In$^2$Test approach, which is able to guide testing activities based on inspection defect data that is available early. With this approach, effort for testing can be allocated more efficiently or more defects can be found. One major question is whether such inspection defect data are a suitable predictor of defect-prone parts. Therefore, we conducted two case studies that first compared inspection with product metrics, and then combined them in order to evaluate their potential.

The remainder of this article is structured as follows: Section 2 presents an overview of established product and process metrics used for prioritizing testing, and introduces the In$^2$Test approach. Section 3 presents experiences from two case studies where inspection and product metrics were compared. Finally, Section 4 concludes the article and gives an outlook on future work.

## II. RELATED WORK

### A. Focusing and Predictions based on Metrics

One well-established approach for focusing quality assurance activities is the prediction of defect distributions based on metrics. One of the earliest metrics, the famous cyclomatic complexity, was introduced by McCabe in 1976; thereafter, a plethora of metrics have been proposed, analyzed, and evaluated. Through a literature survey aimed at obtaining an overview of the most prominent metrics used for predictions, 28 publications covering process and product metrics have been analyzed. Seventeen of these articles covered product metrics, three covered process metrics, and eight covered both types of metrics.

A total of 76 distinct metrics were found of which many were evaluated, split up into four categories. The first major split is the differentiation into process and product metrics. Eleven process metrics and 65 product metrics could be identified. Three subcategories were used to divide the product metrics. The first subcategory consists of 31 different source code metrics. These metrics describe general characteristics of source code and can be applied to any programming language, e.g., class length measured in lines of code. The second subcategory consists of 28 object-oriented metrics. Strictly speaking, these metrics are also source code metrics. However, they make explicit use of object-oriented concepts such as generalization or specialization and therefore cannot be applied to all programming languages. A typical example of an object-oriented metric is 'inheritance coupling', which measures the number of parent classes to which a given class is coupled. The third subcategory, consisting of six further object-oriented metrics, is the well-established metric set introduced by Chidamber and Kemerer called the CK metric suite [5]. The CK metric set consists solely of object-oriented metrics; however, a distinction between the CK metric set and other object-oriented metrics is reasonable, as the CK metric set has been empirically validated many times by various researchers in different contexts.

Certain metrics are more prominent in the literature than others. In order to get an idea of the practical potential of individual metrics, we analyzed how often they had been empirically validated in the 28 articles found. Figure 1 shows an overview of product and process metrics that were evaluated at least six times in the set of articles found.

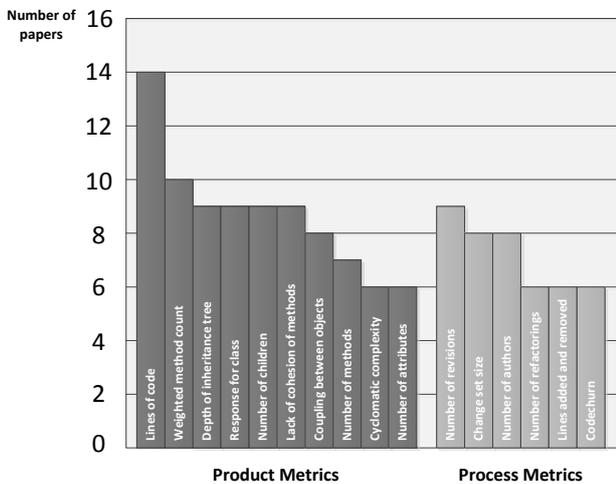

**Figure 1: Number of papers per metric that present evaluation results**

Even though metrics such as 'lines of code' are evaluated often, the results vary significantly. For instance, Gyimóthy et al. stated that large classes are more defect-prone [7], whereas Fenton and Ohlsson conclude that smaller classes are not less likely to be defect-prone than larger classes [8]. One main difference between the two case studies is the different contexts they were conducted in. Consequently, there is no single set of metrics that fits all project contexts [6], and a metric set that fits best in a new context has to be identified before it can be applied to conduct predictions. Different characteristics of a project (e.g., programmer experience, manpower) can have a major influence on the predictive quality of metrics. Furthermore, most of the metrics cannot be applied before code is written. Defect data that might be available earlier are rarely used for focusing subsequent quality assurance activities; e.g., inspection and testing activities are often used in isolation, and inspection metrics are usually not considered for predicting defect-proneness in order to focus testing.

*B. The In²Test Approach*

The **in**tegrated **in**spection and **test**ing approach In²Test was developed to allow using inspection defect data systematically for the prediction of defect-prone parts (e.g., code classes, modules) and defect types for guiding testing activities [3], [4]. The main idea of In²Test is to predict those parts for testing that are expected to be most defect-prone, or to predict those defect types that are expected to show up during testing activities, and to explicitly consider inspection defect data (and optionally further metrics). Two different possibilities exist for prioritization: the first is one-stage prioritization, meaning that only defect-prone parts or defect types are prioritized; the second is two-stage prioritization, meaning that parts of the system under test that are expected to be defect-prone are prioritized first, followed by prioritization of defect types that should be focused on in the prioritized parts. In this article, we consider only prioritization of defect-prone parts.

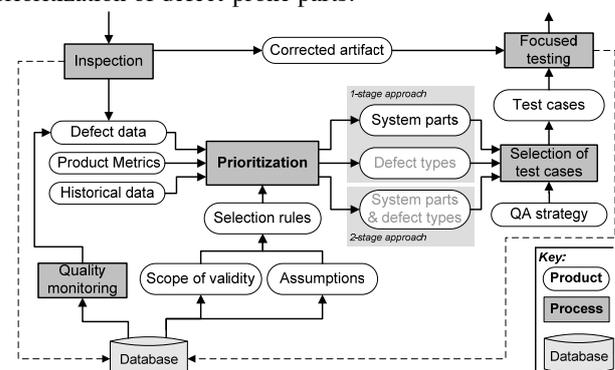

**Figure 2: In²Test approach**

In order to be able to focus testing activities based on inspection defect data, relationships between defects found in the inspection and defects to be found with testing have to be defined. Such knowledge is often not available. For this reason, assumptions need to be defined explicitly. Furthermore, assumptions are often too coarse-grained to be applied. Consequently, refined selection rules have to be derived to make the assumptions operational. For example, one assumption for the In²Test approach can be a Pareto distribution of defects, i.e., parts of a system where a large number of defects are found with inspections indicate that more defects are expected to be found with testing. In this case, it has to be clarified what "large number of defects" means in a concrete environment, i.e., a concrete metric and thresholds have to be defined. Selection rules to be chosen depend on the available and analyzed data from the concrete context. Optionally, product metrics and historical data can be combined with inspection defect data in order to improve the prioritization.

Assumptions are only valid within a certain scope of validity. Therefore, the concrete context has to be gathered, and a significance level has to be considered for each assumption (if an assumption is applied successfully the first time, it gets a significance level of 1, which is increased each time the assumption is evaluated as valid in a quality assurance run).

After prioritization has been completed, test cases have to be derived or selected. The number of test cases used must fit the overall quality assurance strategy and the way prioritization is performed. For example, one strategy might be that only the top-prioritized parts of a system and defect types should be tested, omitting the remaining ones and thus, saving testing effort. Another strategy might be to focus most of the available test effort on the prioritized parts of a system, and to test the remaining parts with little effort. Consequently, the aim is to improve the defect detection ability (i.e., effectiveness). Finally, the prioritized testing activity is performed accordingly, and gathered data (e.g., defect data, validity of assumptions) are stored in a database.

### III. EXPERIENCES

In this section, we describe experiences regarding the performance of various product and inspection metrics for optimizing testing from two different contexts. The first one builds upon the results from a previous case study and extends these analyses. The second one introduces early experiences from an industrial environment.

#### A. Case Study 1

*1) Goal*

The main goal of this study was to evaluate the performance of certain well-established product metrics and inspection defect metrics that are able to focus testing activities. In two earlier case studies [3], [4], the In$^2$Test approach had been evaluated with regard to its feasibility and effort improvement potential. For this, defect and effort numbers gathered during two quality assurance runs, during each of which inspection and testing activities were conducted, were analyzed and compared to each other. However, only a small number of assumptions and selection rules were applied in this study, and no explicit comparison of inspection defect metrics to pure product metrics was conducted. Therefore, in order to compare the integrated inspection and testing approach In$^2$Test with established approaches using product metrics for focusing testing activities, the following two research question were derived:

Research Question 1 (RQ1): Which assumptions and selection rules that consider various inspection and product metrics lead to the best prioritizations of defect-prone code classes?

Research Question 2 (RQ2): Which assumptions and selection rules that consider various inspection and product metrics are stable across several quality assurance runs in a given environment, i.e., which assumptions and selection rules turned out to be most effective during a trend analysis?

*2) Context*

A case study in which two quality assurance (QA) runs were conducted in the same environment in order to analyze the In$^2$Test approach formed the basis for the following analysis.

The artifact to be checked was a Java prototype tool called JSeq, which had mainly been developed by one developer. JSeq supports practitioners in performing sequence-based specifications. At the time of the case study, it consisted of 76 classes, over 650 methods, and about 8,500 lines of code (LoC). The critical code parts were inspected. In the first QA run, these comprised four classes with a total of about 1,000 LoC. In the second run, four classes of about 2,400 LoC were inspected. Due to continuous development of the tool, the inspected code classes were completely different between the two QA runs.

In the first run, one inspector had very good inspection knowledge, but only limited programming experience, whereas the remaining three inspectors were mainly testers or developers with some inspection knowledge, but high programming experience. In the second run, one developer was replaced by an experienced inspector. The testing activity was performed by one developer who was not involved in the inspection.

In the first run, only one assumption considering inspection defect data was applied in order to check the general applicability of the integrated approach. Three assumptions applied in the second run considered only inspection defect data, the combination of inspection defect data and size, and the combination of inspection defect data and complexity. These initially stated assumptions were defined in a group session, and general empirical evidence for each one was found in the literature. A set of derived selection rules that were analyzed in the second run showed an effort improvement for test execution (including test definition) of between 6 and 34 percent.

*3) Design*

In order to perform a detailed analysis of product and inspection metrics for focusing testing activities, we used data from the original study, which had analyzed the In$^2$Test approach during two QA runs [4]. In the study, a code inspection was conducted first each time, followed by quality monitoring of the inspection results, prioritization of the code classes based on the inspection results and further metrics, and application of a unit test. Afterwards, a retrospective analysis of the suitability of the initially defined assumptions and selection rules was conducted. The analysis in this study considers assumptions and selection rules focusing on defect-prone code classes, i.e., no prioritization of defect types is considered here.

In order to be able to perform a comprehensive comparison of different inspection and product metrics and their combinations, the initial set of assumptions and selection rules was heavily extended (i.e., tripled) in a systematic manner. Furthermore, an evaluation scheme had to be defined. For this, we used both a broad- and a fine-grained scale. First, each selection rule was assessed as being either effective or ineffective. Effective in this context means that a selection rule prioritizes all code classes that contain defects, i.e., no defect is omitted by the selection of a subset of all code classes used for testing. On the other hand, ineffective means that a selection rule does not prioritize all or even no code classes that are defect-prone. Such a classification of selection rules presents an initial impression of the performance of the selection rules and the corresponding assumptions, and gives an idea of which prioritizations of code classes can save effort at the same quality level.

Besides this coarse-grained assessment, a more fine-grained one is able to further distinguish the quality of the selection rules; Figure 3 presents concrete examples. Considering effective selection rules first, two categories can be distinguished: A selection rule prioritizes all code classes in which test defects are found, and code classes in which no defects are found during testing are not prioritized (category A). Furthermore, an effective selection rule is classified into category B if all code classes in which test defects are found are prioritized, but also code classes in which no defects are found are prioritized. A corresponding selection rule in the above example would be "Prioritize code classes for testing that contain more than eight defects based on the inspection defect data".

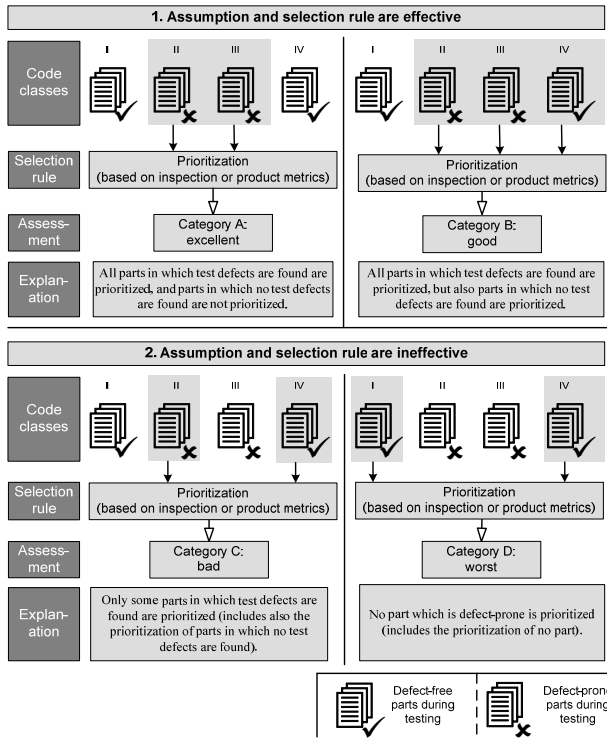

Figure 3: Four quality categories

With respect to ineffective selection rules, two cases are possible. A selection rule is classified into category C if only some code classes in which test defects are found are prioritized, i.e., if some defects are found, but others are overlooked. A combination of such selection rules might improve the prioritization of defect-prone code classes and should therefore be further analyzed during subsequent QA runs. The last category D comprises selection rules that do not prioritize any code classes that turned out to be defect-prone during testing.

Besides the individual analysis of assumptions and selection rules during each of the two QA runs, a trend analysis was performed in order to check which of the selection rules were suited best across both QA runs.

*4) Execution*

In the original study, only one, respectively 32, selection rules were used during the evaluations. The objective in this article is to provide a more comprehensive evaluation of assumptions and selection rules that are valid in the given context, with explicit comparison of inspection and product metrics. All necessary information had already been gathered during the two QA runs (e.g., defect information for each code class, two size and one complexity metrics).

In this study, we defined fourteen different assumptions, two considering only inspection metrics, four considering product metrics, and eight combining inspection metrics and product metrics. For example, with respect to inspection metrics, the following two assumptions are reasonable:

A.I. Parts of the code where a *large* number of inspection defects are found indicate more defects to be found with testing.
A.II. Parts of the code where a *low* number of inspection defects are found indicate more defects to be found with testing.

For each of these assumptions, detailed selection rules were derived systematically, resulting in an overall number of 118 selection rules. Table 1 shows the calculation. One example of a selection rule for A.I is: "Focus testing on those code classes with *large defect density* considering *all inspection defects*." Another example for A.III is: "Focus testing on *large* code classes." A third example for A.XII is: "Focus testing on code classes with *large defect content* and *low complexity* considering *high severity inspection defects* and *low McCabe complexity*."

Table 1: Calculation numbers of selection rules

| Assumptions | | Selection | Metrics one | Metrics two | # |
|---|---|---|---|---|---|
| I | inspection defect data | 2 x | 2 x | 4 | = 16 |
| II | | large / small | defect content / defect density | all defects / high severity defects / med. severity defects / low severity defects | |
| III | size | 2 x | 1 | | = 2 |
| | | large / small | class length | | |
| IV | size | 2 x | 1 | | = 2 |
| | | large / small | method length | | |
| V | complexity | 2 x | 1 | | = 2 |
| VI | | high / low | McCabe complexity | | |
| VII | inspection defect data + size | 4 x | 2 x | 4 | = 32 |
| VIII | | large + large / large + small / small + large / small + small | defect content + class length / defect density + class length | all defects + LoC / high severity defects + LoC / med. severity defects + LoC / low severity defects + LoC | |
| IX | inspection defect data + size | 4 x | 2 x | 4 | = 32 |
| X | | large + large / large + small / small + large / small + small | defect content + method length / defect density + method length | all defects + LoC / high severity defects + LoC / med. severity defects + LoC / low severity defects + LoC | |
| XI | inspection defect data + complexity | 4 x | 2 x | 4 | = 32 |
| XII | | large + high / large + low / small + high / small + low | defect content + McCabe / defect density + McCabe | all defects + McCabe / high severity defects + McCabe / med. severity defecty + McCabe / low severity defects + McCabe | |
| XIII | | | | | |
| XIV | | | | | |
| | | | | Sum: | 118 |

A definition of the general selection criteria "low", "large", etc. was done context-specifically during the original case study, so that all of these selection rules could be applied with respect to the gathered data during the two QA runs in order to allow assessing and comparing them. To give an idea of the defect data and product metrics, Table 2 shows an excerpt (we refer to [4] for the remaining data).

Table 2: Excerpt of metrics

| | QA run 1 | | | | QA run 2 | | | |
|---|---|---|---|---|---|---|---|---|
| Code classes | I | II | III | IV | V | VI | VII | VIII |
| Inspection defects | 26 | 6 | 27 | 8 | 14 | 40 | 39 | 7 |
| Test defects | 3 | 0 | 4 | 0 | 0 | 0 | 6 | 0 |
| Class length | 469 | 37 | 275 | 243 | 231 | 1364 | 701 | 115 |
| Mean method length | 4 | 9 | 7 | 177 | 3 | 14 | 8 | 7 |
| McCabe complexity | 2 | 5 | 2 | 44 | 2 | 4 | 3 | 2 |

*5) Results*

**QA run 1:** In order to answer RQ1, we first analyzed the 118 selection rules with respect to the first QA run. Nineteen selection rules turned out to be effective, and consequently, ninety-nine were ineffective. This is not surprising as we analyzed a large number of rules. Figure 4 gives an overview with respect to the four different categories.

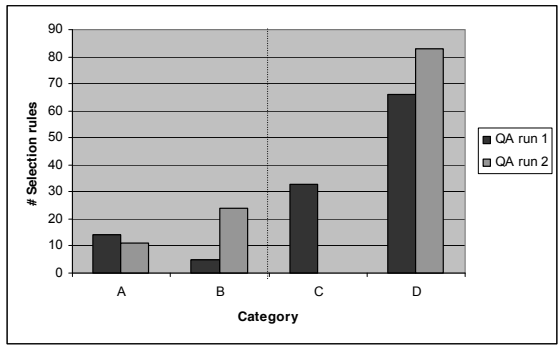

**Figure 4: Assessment result of selection rules**

In our context, the best selection rules (i.e., category A) were those that use large defect content alone or combine this with large class length, small method length, or low complexity. Thus, a Pareto distribution could be confirmed. Rules considering large class length or small method length led to category B. Defect density was a bad predictor for defect-proneness in our context (category C). This means that certain inspection metrics alone and combined with traditional product metrics led to the best selections of defect-prone code classes here, and product metrics alone led to suitable predictions but not to the most efficient ones.

In conclusion, assumptions considering large numbers of inspection defects and low complexity were appropriate. With respect to size, it depends on the concrete size metric. Corresponding combinations also led to suitable selections.

**QA run 2:** Next, we analyzed the 118 selection rules with respect to the second QA run. Thirty-five selection rules were rated as effective. The number for category B increased, which is not surprising due to the fact that only one defect-prone code class was found during testing, and many selection rules select more than one code class. No category C selection rule was found because no subset of one defect-prone code class can exist. However, the general trend of A+B and C+D is comparable to the first QA run.

Again, large defect content alone and the combination with large class length or small method length led to the best selections of code classes. However, large defect density led to much better results in the second QA run. Furthermore, high complexity alone and combined with large defect content and defect density led to suitable results (instead of low complexity as in the first QA run). Selection rules considering large class length or low method length were again evaluated as category B.

Consequently, the Pareto distribution could be confirmed again. While the two size metrics showed similar results compared to the first QA run, namely being effective predictors of defect-proneness while not being most efficient, complexity behaved inconsistently.

Moreover, a combination of inspection and product metrics for focusing testing activities showed the best results in our context (category A). A large number of selection rules were found that led to ineffective results and that are of little relevance for future QA runs in the given context.

In conclusion, assumptions considering large numbers of inspection defects and high complexity were appropriate. With respect to size, it depends on the concrete size metric again. Corresponding combinations also led to suitable selections.

**Trend analysis:** In order to answer RQ2, we analyzed which selection rules were effective, respectively ineffective, with respect to both QA runs. Figure 5 presents an overview of the results.

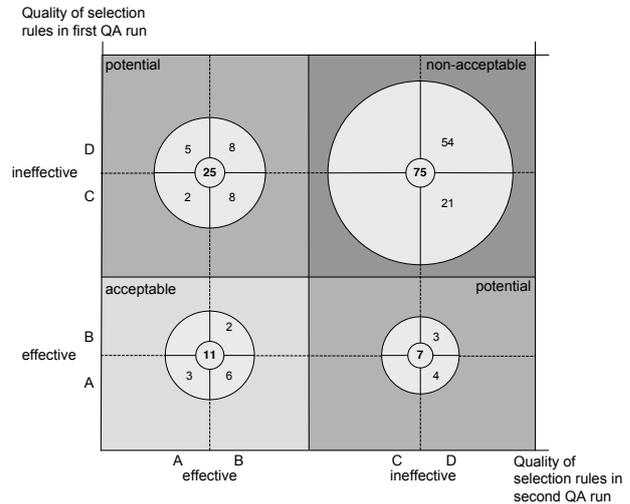

**Figure 5: Trend analysis of selection rules**

First of all, the general classification of selection rules into effective and ineffective ones over the two QA runs revealed that only about 10% of the selection rules were effective in both runs (acceptable box). These eleven rules are promising candidates in the given environment for a highly effective prediction of defect-prone parts. About 30% showed good results in one run, both bad in the other (potential box). Those should be further analyzed, e.g., whether certain context factors can explain those differences, and how they behave in subsequent QA runs. The remaining 75 selection rules showed ineffective results in both QA runs and are thus of little interest for future runs (non-acceptable box). The high number of such classified rules is not surprising, as we compared a large number of selection rules.

With respect to the acceptable selection rules, three were classified into category A in both runs, i.e., these selection rules selected exactly the defect-prone code classes for testing based on large defect content (all, medium, and low severity) combined with small method length. Six more selection rules showed very promising results, considering only large defect content, and large defect content combined with large class length. Two more selection rules were twice categorized as 'B', namely large class length and small method length. This means that inspection defect data alone (in terms of defect content) and inspection defect data combined with certain product metrics led to the best prioritizations in our context. Furthermore, two size metrics led to appropriate selections of code classes, though not to the most efficient prioritizations. This also holds for the corresponding assumptions. Table 3 lists those metrics that led to the best prioritizations of code classes containing defects found during testing.

**Table 3: Best metrics for prediction defect-proneness in the given context during two QA runs**

| Quality | Metric |
|---|---|
| AA | High inspection defect content and low method length (all, low, and medium severity) |
| AB | High inspection defect content (all, low, and medium severity) |
| | High inspection defect content and high class length (all, low, and medium severity) |
| BB | High class length |
| | Low method length |

A lot of selection rules considering high defect density alone or in combination with the aforementioned product metrics led to category C prioritizations and might lead to more suitable results in future QA runs. Furthermore, selection rules using complexity led to inconsistent selections in our context. While in the first QA run, low complex code classes were more defect-prone, this changed in the second QA run, and high complex code classes tended to be more defect-prone. One explanation is that the first QA run was performed when the software was still not very complex, and thus, such parts also contained defects.

*6) Threats to Validity*

Conclusion validity: The presented results are only based on two QA runs in a given context and therefore, a lot more evaluations in the same and in different contexts are needed before they can be generalized. However, first positive trends could be identified indicating that inspection defect results lead to good prioritization of defect-prone parts for testing, and that a combination with established product metrics might even improve such a prioritization.

Construction validity: A set of assumptions and selection rules was derived systematically for our analysis. However, a lot of additional ones might exist, and a comparison with more product metrics would strengthen such an analysis.

Internal validity: The evaluation of selection rules might have been done differently by other QA engineers who might have defined thresholds for 'large' or 'small' in a different way. However, the thresholds were discussed in a team of quality assurance engineers to reduce this threat.

External validity: The software under inspection and test during the QA runs was rather small, and only a small set of code classes was considered for the prioritizations. Furthermore, the performance of the applied assumptions and selection rules is initially only valid in the given context and cannot be generalized, as each such rule has to be evaluated again in each new context. However, we presented an initial set of assumptions and selection rules that showed promising results and which are consistent with existing empirical knowledge (such as the Pareto distribution); thus, our rules might serve as a starting point for evaluations in new environments, and can be enhanced by additional or alternative rules.

B. *Case Study 2*

*1) Goal*

Based on the experiences from the first case study, we started an evaluation of the In$^2$Test approach in an industrial environment. The main goal of the second case study was to evaluate which assumptions and selection rules lead to the most effective prioritizations of parts for testing. Once again, different inspection and product metrics used to select defect-prone parts were compared. Similar to RQ1, RQ3 is stated as follows:

Research Question 3 (RQ3): Which assumptions and selection that consider various inspection and product metrics rules lead to the best prioritizations of defect-prone modules?

*2) Context*

The analyzed organizational unit has been developing software for deeply embedded systems, mainly automotive, for over 20 years. Currently, Vector uses a product family approach for development with three levels of variation: the product as the full superset of all features, the program as the first derived level of variation for a specific customer platform, and the delivery as second level of variation for a concrete microprocessor and compiler. The features are implemented in the form of components.

As a consequence of the sensitive context, several activities are performed to ensure the quality of the software. On the code level, code inspections are conducted on all released source code. Testing is done on all elements of the product family: the product itself, programs, deliveries, and components. The defect data and all feedback from customers are stored in the form of change requests. Even though the development model of the product family has changed over time, test and inspection data are available spanning more than ten years.

The software currently consists of about 140 modules, each comprising a set of code classes. The size of a module varies between 120 and 14,000 statement lines of code.

*3) Design*

In order to evaluate the In$^2$Test approach and a set of different assumptions and selection rules, a retrospective design was chosen again. For this, existing inspection data had to be collected first, as all such data were documented across several change request documents. For our analysis, we concentrated on a subset of twelve of the available modules (about 10%), which we chose randomly. The reason is that size, complexity, age, and number of deliveries are highly heterogeneous. The first question was whether we could find basic assumptions that work independent of any component classification. For each module, a different number of change requests (and thus inspection data) existed. The oldest defect data that was considered was from 2007.

With respect to test data, we considered all defects found during various kinds of testing performed subsequent to the code inspection.

Furthermore, two code metrics (i.e., product metrics) were considered and calculated for the corresponding modules: size in lines of code and waste per line, which express the stability and sustainability of the developed code. A high value implies that many parts are changed or thrown away over time.

For the analysis of the approach, we defined ten assumptions, based upon the available data from the context and our experience from the first case study. Furthermore, due to the defect distribution, a categorization of assumptions into effective and ineffective turned out not to be useful because each module contained at least some minor problems over the considered timeframe, and all assumptions would therefore have been classified into the detailed category C (which would have made them impossible to compare). Therefore, we defined four selection rules for each assumption that, based on the general assumption, select, e.g. the three most defect-prone modules based on the inspection results, respectively the three largest modules based on the product metrics. We continued this with the top-5, the top-8, and the top-10 modules, and evaluated how many test defects had been found by such selections, i.e., we derived a sorted list instead of defining a set based on a hard threshold. With such an analysis, a baseline of appropriate assumptions could be gathered for the given context.

*4) Execution*

We first presented the In$^2$Test approach to Vector and discussed the expected benefit in their environment. As improving quality assurance is a major goal for Vector, we decided to evaluate the approach in a retrospective design based on the historical inspection and test defect data. We first gathered inspection data from several change request documents for the randomly chosen code classes, which took several days. The test data could be extracted easily from a defect tracking system. However, chronology sequence and relations between the inspection and test data were difficult to extract. Therefore, we decided to use the existing data to draw a baseline from which we assessed each assumption.

Ten assumptions were derived: four considering inspection metrics, four considering product metrics, and two combining inspection and product metrics. With respect to those using inspection defect data, a Pareto distribution was assumed, and different representations of the inspection data were used:

1. All inspection defect data for a module.
2. Like 1, but scaled (not all modules were inspected 100%, but the rate was given and could be used to estimate the inspection defect numbers if 100% had been inspected).
3. All inspection defect data without counting inspection comments.
4. Like 3, but scaled.

With respect to assumptions considering product metrics, modules of small and large size, and modules of small and large waste per line are taken.

Four selection rules were derived for each of the assumptions following the structure shown in Figure 6. One concrete example: "A1: Focus testing on the top-3 defect-prone code classes based on all inspection defect data."

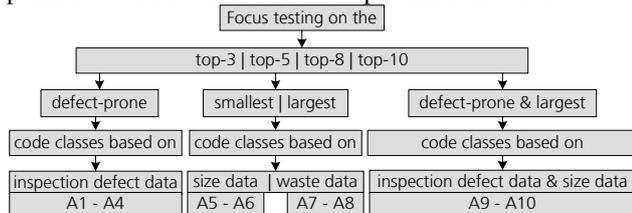

**Figure 6: Structure of applied selection rules**

Based on the available defect data and the product metrics, all assumptions and selection rules were applied in a retrospective manner in order to evaluate their validity.

*5) Results*

Figure 7 presents an overview of the number of found test defects with respect to the ten assumptions. For each assumption, four different selection rules were evaluated. For example, assumption A1 assumes that a large number of the test defects were found in those modules where most inspection defects had been found before. With respect to the first selection rule, which considers the top-3 defect-prone inspection modules, about 30% of all test defects had been found. Considering the top-5 modules, more than 80% of all test defects had been found. This means that considering about 40% of all defect-prone modules based on the inspection defect data was sufficient for finding more than 80% of all defects during testing. Focusing on the top-8 defect-prone modules during inspections, more than 90% of all test defects were found. The top-10 do not further improve prioritization when using assumption A1.

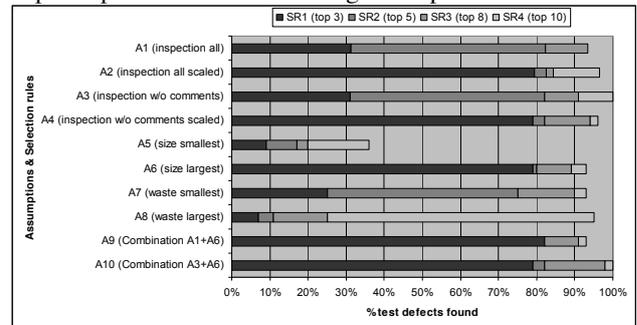

**Figure 7: Results of different assumptions**

Without initially considering the combined assumptions A9 and A10, assumptions A2, A4, and A6 led to the best results with respect to the top-3 modules, i.e., those selection rules need only 25% of the defect-prone modules to focus testing in such a way that about 80% of the remaining defects are found. Two of these assumptions consider inspection results, one assumption considers size. With respect to the top-5, all four inspection assumptions led to suitable predictions (more than 80% of test defects were found). Only one selection rule (from assumption A3) revealed all defect-prone modules with the top-10 selection (and thus, can be classified as "effective" in the sense of the first case study). However, almost no selection rule was able to prioritize all modules containing defects in that context due to the fact that almost every module contained defects, and the selection should identify a subset of modules in order to save effort while maintaining the same quality. Selection rules considering small size and large waste did not lead to suitable results.

Finally, we combined inspection and product metrics and also calculated the number of defects found. For example, when combining A1 and A6 and calculating the top-3 value, we counted each module selected by one of the two assumptions, which resulted in two modules that were selected by either A1 and A6, and two modules that were selected by only one of the two assumptions. The resulting effectiveness value was 82%. Surprisingly, the top-5 calculation did not improve the value significantly, but the modules selected for the top-8 did. This indicates that one should focus on modules that fulfill the combined assumption best (i.e., top-3 focus) to find about 80% of the test defects, and on those modules in the top-5 to top-8 range of the combined assumption (by skipping those in the top-3 to top-5 range) to find another 15% of the remaining test defects in a given context.

Two main conclusions can be drawn: The assumptions that considered inspection defect data (1) led to suitable predictions, and (2) are of similar effectiveness as selection rules using size metrics. Though not all defects were found by most of the selections due to the long timeframe that was considered, the most critical parts could be identified by these assumptions. Because the modules were selected randomly, our objective is to further investigate whether these product metrics and inspection metrics behave similar with respect to a broader dataset, and whether the prediction can be further improved when combining them.

*6) Threats to Validity*

Conclusion validity: The presented results are only based on the analysis of a subset of all available modules from the context. Therefore, a larger analysis is still necessary. However, the initial results substantiate similar trends from different environments.

Construction validity: It is possible to evaluate a set of additional assumptions and selection rules, which might lead to further conclusions. However, we chose such metrics in our assumptions and rules that were already available in the given context.

Internal validity: A certain degree of inaccuracy is often a fact with respect to historical defect data. However, the absolute number of documented defects was large enough to compensate for that to a certain extent.

External validity: First, it could be verified in the given context that large modules tend to be more defect-prone, which is consistent with evaluations from different contexts. Furthermore, the prioritization for testing based on the inspection metrics led to similar results for some assumptions, or even slight improvements compared to the product metrics. Though this is a positive trend in this environment, conclusions with respect to other environments have to be drawn with caution, as other assumptions might lead to good selections in different contexts. The assumptions used in this study can serve as a starting point for such evaluations.

## IV. Summary and Outlook

The prediction of defect-prone software parts is one way of improving quality assurance activities. Usually, product and process metrics are used for such a prioritization. However, inspection defect data is usually not considered. Therefore, we propose the In$^2$Test approach, which is able to use inspection defect data alone or in combination with such product and process metrics. This approach is not intended to substitute existing approaches, but rather complement them in order to further support the planning of quality assurance activities.

In order to evaluate the suitability of inspection and product metrics for predictions of defect-prone parts, we conducted two studies and compared different metrics. We could show that inspection defect data were an appropriate predictor in those two contexts, which was further improved when inspection metrics were combined with certain product metrics. Our results represent promising, but initial results, and more empirical studies are necessary before generalizing our conclusions.

We are planning to continue evaluations of the In$^2$Test approach in order to substantiate our findings and to find more relationships between inspection and test defect data. Furthermore, a lot of additional product metrics exist that can be compared and combined with inspection defect data. Further evaluation of selection rules can substantiate our findings, e.g., by calculating precision and recall values. In addition, results from requirements and design inspections might help to better focus subsequent quality assurance activities. Finally, the approach could also be extended in such a way that test data might be used for improving the inspection, as the empirical concepts for evaluating assumptions and selection rules are generalizable.

ACKNOWLEDGMENT

This work has been funded by the Stiftung Rheinland-Pfalz für Innovation project "Qualitäts-KIT" (grant: 925) and the ARTEMIS project "MBAT" (grant: 269335). We would also like to thank Sonnhild Namingha for proofreading.